\documentclass[journal,comsoc]{IEEEtran}
\usepackage[T1]{fontenc}% optional T1 font encoding
\usepackage{xcolor}
\usepackage{fancyhdr}
\usepackage{tgschola}
\usepackage{lastpage}
\usepackage{amsthm,amssymb,amsmath,mathtools,fixmath}
\usepackage{amsbsy}
\usepackage{moreverb}
\usepackage{amsfonts}
\usepackage{graphicx}
\graphicspath{{SimulationFigures/}}
\usepackage{tikz}
\usepackage{listings}
\usepackage{color}
\usepackage{latexsym}
\usepackage{subfigure}
\usepackage{animate}
\usepackage{nopageno}
\colorlet{linkequation}{blue}
\usepackage[linesnumbered,ruled]{algorithm2e}
\usepackage{soul}
\usepackage{fancyhdr}
\usepackage{kantlipsum}

\fancypagestyle{plain}{
  \fancyhf{}
  \fancyhead[C]{Conference on \LaTeX}     %% C or L or R.
  \fancyfoot[L]{This is a notice}%                        %% C or L or R.

}
\usepackage{eso-pic}

\def\BibTeX{{\rm B\kern-.05em{\sc i\kern-.025em b}\kern-.08em T\kern-.1667em\lower.7ex\hbox{E}\kern-.125emX}}

\usepackage{geometry}
\usepackage{lipsum}% http://ctan.org/pkg/lipsum

\makeatletter
\def\BState{\State\hskip-\ALG@thistlm}
\makeatother

\usepackage{cite}
\ifCLASSINFOpdf
\else
\fi
\usepackage{amsmath}
\usepackage[cmintegrals]{newtxmath}
\begin{document}
%\title{Matching Theory for User Association in \\mmWave-Enabled Heterogeneous Networks}
% \title{A Realistic Interference Model for \\Millimeter Wave Cellular Networks}
\AddToShipoutPictureBG*{%
  \AtPageUpperLeft{%
    \setlength\unitlength{1in}%
    \hspace*{\dimexpr0.5\paperwidth\relax}%%  change \dimexpr0.5\paperwidth\relax appropriately
    \makebox(0,-0.75)[c]{\small A. Alizadeh, M. Vu, and T. S. Rappaport, ``A Study of Interference Distributions in Millimeter Wave Cellular Networks,''}
    \makebox(0,-1)[c]{\small Accepted for publication in IEEE Conference on Microwaves, Communications, Antennas and Electronic Systems, Nov. 2019.}  %
}}
\title{A Study of Interference Distributions in \\Millimeter Wave Cellular Networks}

\author{\IEEEauthorblockN{Alireza Alizadeh\IEEEauthorrefmark{1}, Mai Vu\IEEEauthorrefmark{1}, and Theodore S. Rappaport\IEEEauthorrefmark{2} \\}
\IEEEauthorblockA{\IEEEauthorrefmark{1}Department of Electrical and Computer Engineering,
Tufts University, Medford, MA, USA\\
\IEEEauthorrefmark{2}NYU WIRELESS, Tandon School of Engineering, New York University, Brooklyn, NY, USA\\
Email: \IEEEauthorrefmark{1}\{alireza.alizadeh, mai.vu\}@tufts.edu, \IEEEauthorrefmark{2}tsr@nyu.edu}
%\vspace*{-2em}
%\and
%\IEEEauthorblockN{Mai Vu}
%\IEEEauthorblockA{Department of Electrical and Computer Engineering\\
%Tufts University\\
%Medford, USA\\
%Email: mai.vu@tufts.edu}
\thanks{The work in this paper is supported in part by the National Science Foundation under CNS Grants 1908552 and 1909206.}
}

%\author{Alireza~Alizadeh, and Mai~Vu% <-this % stops a space
%\thanks{M. Shell was with the Departmentof Electrical and Computer Engineering, Georgia Institute of Technology, Atlanta, GA, 30332 USA e-mail: (see http://www.michaelshell.org/contact.html).}% <-this % stops a space


\markboth{}%
{Shell \MakeLowercase{\textit{et al.}}: }



\maketitle
\begin{abstract}
We study the distribution of the interference power in a millimeter wave (mmWave) cellular network. Such interference is random and highly dependent on the employed transmission technique, as well as the varying channel conditions and the varying association between users and base stations.
Traditional networks at lower frequencies usually employ omnidirectional transmission which creates an (almost) equal amount of interference in any direction. MmWave networks, however, must employ directional beamforming transmission in order to compensate for the high path loss in mmWave frequency bands. These directional transmissions drastically change the network interference structure. 
We examine the interference power distributions in an mmWave network employing beamforming transmission under different user association schemes, and contrast with those under omnidirectional transmission. Numerical results using an analytical mmWave channel model and \textcolor{black}{a measurement-based channel generator, NYUSIM,} show that \textcolor{black}{beamforming not only} reduces the amount of strong interference and hence significantly enhances network throughput, but also user association can considerably alter network interference and throughput structures.
\end{abstract}

%\begin{IEEEkeywords}
%Matching theory, user association, deferred acceptance, early acceptance, 5G cellular HetNets.
%\end{IEEEkeywords}

% \fi
\IEEEpeerreviewmaketitle

%%%%%%%%%%%%%%%%%%%%%%%%%%%%%%%%%%%%%%%%%%%%%%%%%%%%%%%%%%%%%%%%%%%%%%%%%%%%%%%
%%%%%%%%%%%%%%%%%%%%%%%%%%%%%%%%%%%%%%%%%%%%%%%%%%%%%%%%%%%%%%%%%%%%%%%%%%%%%%%

\section{Introduction}
\thispagestyle{empty}
Millimeter wave (mmWave) technology is a key enabler of future 5G cellular networks. To accommodate for the shorter signal range, mmWave-enabled networks will have a high density of base stations (BSs). In such a dense network, interference is an important factor affecting network performance, and techniques for interference mitigation or exploitation require an accurate understanding of the interference characteristics. This interference is random depending on the transmission technique, distributions of user equipments (UEs) and BSs, the varying channel conditions, and the varying associations between UEs and BSs. All these factors affect how the interference power at a typical receiver can vary, which is captured by the interference power distribution.

The transmission technique has a strong effect on the interference.  In sub-6 GHz systems, the transmissions are omnidirectional and thus the interference originated from a transmitter propagates approximately equally in all directions. In mmWave systems, however, the transmissions are highly directional because of the use of beamforming \textcolor{black}{at both transmitter and receiver sides } and thus will significantly affect the interference at a typical receiver. In order to 
establish these directional transmissions, we first need to perform user association to establish the connection between specific BSs and UEs based on the channel conditions to maximize a certain network utility.

Max-SINR is a conventional user association scheme in which each UE connects to the BS providing the highest signal to interference and noise ratio (SINR). 
\textcolor{black}{In this approach, however, the BSs which transmit at higher power levels may be overloaded by the surrounding UEs which receive the highest SINR from these BSs.} Load balancing user association schemes have been introduced to resolve this issue by moving the overloading users to lightly loaded base stations to maintain network performance \cite{Andrews,Caire,60GHz,TWC}. Load balancing user association schemes have been proposed for microwave LTE networks \cite{Andrews} and massive MIMO networks \cite{Caire}, and also mmWave WiFi networks at 60 GHz frequency \cite{60GHz}. These works apply omnidirectional transmissions \cite{Andrews,Caire} or consider beamforming transmission but assuming interference is negligible in a short-range WiFi network \cite{60GHz}. This assumption, however, is inaccurate in a cellular mmWave network as the network can transit from a noise-limited regime to an interference-limited regime \cite{Niknam}.
% \cite{NoiseOrInterf,Niknam}.
% M. Rebato et al., “Understanding noise and interference regimes in 5G millimeter-wave cellular networks,” in Proc. European Wireless Conf., May 2016, pp. 1–5.
Moreover, because of beamforming transmissions, the interference structure is highly dependent on user associations in mmWave cellular networks \cite{TWC}. 

In this paper we analyze the realistic interference distribution, called beamforming interference model (BIM), \textcolor{black}{in mmWave systems where singular value decomposition (SVD) beamforming technique is utilized for highly directional transmissions \cite{TWC}}. Since beamforming needs to be performed based on the actual BS-UE connections (associations), the interference structure is also highly dependent on user association. 
We contrast mmWave interference distributions under beamforming transmission with those under omnidirectional transmission as often considered in the literature for lower frequency (e.g. LTE) networks. We further compare the effect of mmWave channel models on the interference distributions by considering 
\textcolor{black}{an analytical channel model (ACM) based on the 3GPP channel model \cite{SS,Nokia} and the measurement-based channel simulator NYUSIM developed from extensive field measurements in New York City \cite{NYUSIM, Rappaport_TVT}.}
%rappaport2017investigation

%%%%%%%%%%%%%%%%%%%%%%%%%%%%%%%%%%%%%%%%%%%%%%%%%%%%%%%%%%%%%%%%%%%%%%%%%%%%%%%
%%%%%%%%%%%%%%%%%%%%%%%%%%%%%%%%%%%%%%%%%%%%%%%%%%%%%%%%%%%%%%%%%%%%%%%%%%%%%%%
\section{System and Channel Models}
%In this section, we introduce our system model as well as millimeter wave channel models.
\subsection{System Model}
We consider a downlink scenario in an mmWave cellular network with $J$ BSs and $K$ UEs. Let $\mathcal{J}$, $\mathcal{K}$ denotes the set of all BSs and all UEs, respectively.   
$M_j$ and $N_k$ are the respective number of antennas at BS $j$ and UE $k$. Each UE $k$ aims to receive $n_k$ data streams (layers) from its serving BS such that $1\leq n_k\leq N_k$, where the upper inequality comes from the fact that the number of data streams for each UE cannot exceed the number of its  antennas.
Each BS can have a quota $Q_j$ which is determined based on its available data streams. 
We define the total number of downlink data streams sent by BS $j$ as
\begin{equation}\label{D_j}
D_j=\sum_{k \in \mathcal{Q}_j}n_k
\end{equation}
where $\mathcal{Q}_j$ is called the \textit{activation set} of BS $j$ which represents the set of active UEs in BS $j$, such that  $\mathcal{Q}_j \subseteq \mathcal{K}$, $|\mathcal{Q}_j(t)|=Q_j(t)\leq K$, and $Q_j$ is the quota of BS $j$. Note that the total number of downlink data streams sent by each BS should be less than or equal to its number of antennas, i.e., $D_j \leq M_j$. 

\subsection{Channel Model}
In the sub-6 GHz band the transmissions are omnidirectional and the entries of Gaussian MIMO channel are i.i.d. complex Gaussian random variables.
In the mmWave band, the transmissions are highly directional and we can not use the simple Gaussian MIMO channel. In this paper, we employ two clustered mmWave channel models which include $C$ time clusters with $L$ subpaths per cluster. 
In Section \ref{NumRes}, we use these channel models to evaluate the performance of our proposed interference model.

\vspace*{.6em}
\subsubsection{MmWave analytical channel model}
The ACM is a clustered channel model defined as  \cite{SS}, \cite{Nokia}
\begin{align}\label{clustered_ch}
H=\frac{1}{\sqrt{CL}}\sum_{c=1}^{C}\sum_{l=1}^{L} \sqrt{\gamma_c}~\mathbf{a}(\phi_{c,l}^{\textrm{UE}},\theta_{c,l}^{\textrm{UE}}) ~\mathbf{a}^*(\phi_{c,l}^{\textrm{BS}},\theta_{c,l}^{\textrm{BS}})
\end{align}
where $\gamma_c$ is the power gain of the $c$th cluster. The parameters $\phi^{\textrm{UE}}$, $\theta^\textrm{UE}$, $\phi^\textrm{BS}$, $\theta^\textrm{BS}$ represent azimuth angle of arrival (AoA), elevation angle of arrival (EoA), azimuth angle of departure (AoD), and elevation angle of departure (EoD), respectively. These parameters are generated randomly based on different distributions and cross correlations as given in \cite[Tables 1-3]{Nokia}. The vector $\mathbf{a}(\phi,\theta)$ is the antenna array response vector, which depends on antenna geometry, including uniform linear array (ULA) or uniform planar array (UPA). In order to enable beamforming in both azimuth and elevation directions (3D beamforming), we use the uniform $U\times V$ planar array given by \cite{SS}
%\frac{1}{\sqrt{S}}
\begin{align}
\mathbf{a}(\phi,\theta)=&\big[ 1, ..., e^{jkd_{\textrm{a}}(u\sin(\phi)\sin(\theta)+v\cos(\theta))}, ...,\nonumber \\  &e^{jkd_{\textrm{a}}((U-1)\sin(\phi)\sin(\theta)+(V-1)\cos(\theta))} \big]^T
\end{align}
where $d_a$ is the distance between antenna elements, and $u\in\{1, ..., U\}$ and $v\in\{1, ..., V\}$ are the indices of antenna elements.

%, and $S=UV$ is the antenna array size.

%Some of the channel parameters are delay spread, azimuth angle of departure (AoD) spread, azimuth angle of arrival (AoA) spread, elevation angle of departure (EoD) spread, elevation angle of arrival (EoA) spread, EoD bias, EoA bias, line of sight (LoS) shadow fading, (non-line of sight) NLoS shadow fading. These parameters are generated 
We consider two link states for each channel, line of sight (LoS) and non-line of sight (NLoS), and use the following probability functions obtained based on the New York City measurements in \cite{RapLetter}
\begin{align}\label{p_LoS/NLoS}
p_{\text{LoS}}(d)&=\Big[\min\Big(\frac{d_\textrm{BP}}{d},1\Big).\Big(1-e^{-\frac{d}{\eta}}\Big)+e^{-\frac{d}{\eta}} \Big]^2\\
p_{\text{NLoS}}(d)&=1-p_{\text{LoS}}(d)
\end{align}
where $d$ is the 3D distance between UE and BS in meters, $d_{\textrm{BP}}$ is the breakpoint distance at which the LoS probability is no longer equal to 1, and $\eta$ is a decay parameter. The obtained values based on measurements for these parameters are $d_{\textrm{BP}}=27$ m and $\eta=71$ m.

%The probability model in () is obtained based on the New York City measurements collected in [3,4 of Rappaport, letter].
Moreover, we use the following path loss model for LoS and NLoS links \cite{RapLetter}%\cite{Nokia}
 % Sigma_SF = 4.9*randn Rappaport's paper Eq. (1) LoS
\begin{align}\label{PL}
PL[\textrm{dB}]=20\log_{10}\Big(\frac{4\pi d_0}{ \lambda}\Big) + 10n \log_{10}\Big(\frac{d}{d_0}\Big) + X_{\sigma_{\textrm{SF}}}
\end{align}
where $\lambda$ is the wavelength, $d_0$ is the reference distance, $n$ is the path loss exponent, and 
$X_{\sigma_{\textrm{SF}}}$ is the log-normal random variable with standard deviation $\sigma_{\textrm{SF}}$ (dB) which describes the shadow fading.
%\begin{align}
%PL(dB)=20\log_{10}\Big(\frac{4\pi d_0}{ \lambda}\Big) + 10n \log_{10}\Big(\frac{d}{d_0}\Big) + \sigma_{\textrm{SF}}
%\end{align}
These parameters vary depending on LoS of NLoS propagation at 28 GHz, the path loss exponents and the shadowing factors are $n_{\textrm{LoS}}=2.1$, $n_{\textrm{NLoS}}=3.4$, $\sigma_{\textrm{SF, LoS}}=3.6$ dB, and $\sigma_{\textrm{SF, NLoS}}=9.7$ dB \cite{RapLetter}.

%These parameters vary depending on LoS of NLoS propagation at 73 GHz, the path loss exponents and the shadowing factors are $n_{\textrm{LoS}}=2$, $n_{\textrm{NLoS}}=3.4$, $\sigma_{\textrm{SF, LoS}}=4.8$ dB, and $\sigma_{\textrm{SF, NLoS}}=7.9$ dB .

In LTE cellular networks, pilot signals are used to estimate channel state information (CSI) at the receiver. Once the CSI is available at the receiver, it can be shared with the transmitter via limited feedback or channel reciprocity. 
However, in 5G dense cellular networks these conventional approaches are inapplicable due to network densification and the limited amount of pilot resources. To address the new challenges emerging from network densification, a promising radio access technology, called cloud radio access network (C-RAN), has been proposed. In this radio access network, CSI at both the transmitter and the receiver can be estimated through new CSI acquisition schemes and shared via C-RAN for centralized signal processing, coordinated beamforming, and resource allocation in 5G new radio \cite{HetNetC-RAN}. In this paper, we assume that the CSI is estimated for the aforementioned applications and we can utilize it for the purpose of user association. 

%In LTE cellular networks, pilot signals are used to estimate channel state information (CSI) at the receiver. Once the CSI is available at the receiver, it can be shared with the transmitter via limited feedback or channel reciprocity. However, in 5G dense cellular networks these conventional approaches are inapplicable due to network densification and the limited amount of pilot resources \cite{User-centricC-RAN}. To address the new challenges emerging from network densification, a promising radio access technology, called cloud radio access network (C-RAN), has been proposed \cite{HetNetC-RAN}. In this radio access network, CSI at both the transmitter and the receiver can be estimated through new CSI acquisition schemes and shared via C-RAN for centralized signal processing, coordinated beamforming, and resource allocation in 5G new radio \cite{Large-ScaleC-RAN}, \cite{5GNR}. In this paper, we assume that the CSI is estimated for the aforementioned applications and we can utilize it for the purpose of user association. 

\subsubsection{Measurement-based NYUSIM channel generator}
% \textbf{Copy form NYUSIM paper \cite{NYUSIM}:}
NYUSIM is a channel generator simulation platform which can generate realistic mmWave channels based upon its input parameters \cite{NYUSIM, Rappaport_TVT}. The models employed in NYUSIM to generate channels are based on extensive measurements at several mmWave frequencies in New York City. The statistical spatial channel model (SSCM) in NYUSIM utilizes time clusters and spatial lobes to model the directional channel impulse response based on AoD/AoA, which have been used successfully in modeling mmWave channels \cite{samimi20163}. Time clusters are composed of subpaths traveling closely in time and arriving from potentially different angular directions. This channel model is motivated by field measurements, which have shown that multiple subpaths within a time cluster can arrive from almost the same directions. 

%The TCSL approach implements a physically-based clustering scheme derived from field observations, and can be used to extract time cluster and spatial lobe statistics for any ray tracing or measurement data sets \cite{samimi20163}.

% It is worth noting that in 3GPP TR 38.900 Release 14 for above 6GHz [32], the number of clusters is unrealistically large. For example, in the UMi street canyon scenario, the number of clusters in the LOS environment is as high as 12, and 19 in the NLOS environment, which is not supported by the real-world measurements at mmWave bands [11]–[14]. In IEEE ICC 2017 Wireless Communications Symposium contrast, in the SSCM upon which NYUSIM is based [13], the number of time clusters ranges from 1 to 6, and the mean number of spatial lobes is about 2 and is upper-bounded by 5, which are obtained from field observations and are much smaller than those in the 3GPP channel model [32]. The impractically large number of clusters in the 3GPP channel model causes a higher rank of simulated mmWave channels, and unrealistic eigen-channel distributions, thereby yielding inaccurate

%%%%%%%%%%%%%%%%%%%%%%%%%%%%%%%%%%%%%%%%%%%%%%%%%%%%%%%%%%%%%%%%%%%%%%%%%%%%%%%
%%%%%%%%%%%%%%%%%%%%%%%%%%%%%%%%%%%%%%%%%%%%%%%%%%%%%%%%%%%%%%%%%%%%%%%%%%%%%%%

\section{Omnidirectional and Directional Transmissions and Associated Interference}
In this subsection, we describe the omnidirectional and directional transmitted signal models and develop a signal model for the interference at a typical receiver.

\subsection{Omnidirectional Transmission}
In a multi-user omnidirectional transmission, the transmitted signal from the antennas of BS $j$ is simply the summation of the intended signals for each user and it can be defined as
\begin{equation}\label{x_j1}
\mathbf{x}_j = \sum_{k\in \mathcal{Q}_j}\mathbf{s}_k
\end{equation}
where $\mathbf{s}_k\in \mathbb{C}^{M_j}$ is the data stream vector intended for UE $k$ composed of mutually uncorrelated zero-mean symbols, with $\mathbb{E}\lbrack \mathbf{s}_k\mathbf{s}_k^*\rbrack = \mathbf{I}_{M_j}$. $\mathbb{E}[\mathbf{x}_j^* \mathbf{x}_j]\leq P_j $ describes the power constraint at BS $j$, where $P_j$ is the transmit power of BS $j$.

The received signal of UE $k$ is given by
\begin{equation}\label{y_k}
\mathbf{y}_k = \sum_{j\in \mathcal{J}}\mathbf{H}_{k,j}\mathbf{x}_j + \mathbf{z}_k
\end{equation}
where $\mathbf{H}_{k,j}\in\mathbb{C}^{N_k\times M_j}$ represents the channel matrix between BS $j$ and UE $k$, and $\mathbf{z}_k\in\mathbb{C}^{N_k}$ is the white Gaussian noise vector at UE $k$, with $\mathbf{z}_k\sim\mathcal{CN}(\mathbf{0},N_0 \mathbf{I}_{N_k})$ and $N_0$ is the noise power spectral density. 

\subsection{Directional Beamforming Transmission}
In a multi-user directional transmission, the transmitted signal from the antennas of BS $j$ can be expressed as
\begin{equation}\label{x_j2}
\mathbf{x}_j = \mathbf{F}_j \mathbf{d}_j = \sum_{k\in \mathcal{Q}_j}\mathbf{F}_{k,j}\mathbf{s}_k
\end{equation}
where $\mathbf{s}_k\in \mathbb{C}^{n_k}$ is the data stream vector intended for UE $k$ composed of mutually uncorrelated zero-mean symbols, with $\mathbb{E}\lbrack \mathbf{s}_k\mathbf{s}_k^*\rbrack = \mathbf{I}_{n_k}
$. The column vector $\mathbf{d}_j\in \mathbb{C}^{D_j}$ represents the vector of data symbols of BS $j$, which is the vertical concatenation of the data stream vectors $\mathbf{s}_k,~k\in\mathcal{Q}_j$, such that $\mathbb{E}\lbrack \mathbf{d}_j\mathbf{d}_j^*\rbrack = \mathbf{I}_{D_j}$.
\textcolor{black}{
Matrix $\mathbf{F}_{k,j}\in\mathbb{C}^{M_j\times n_k}$ is the linear precoder (transmit beamforming matrix) for each UE $k$ associated with BS $j$ which separates user data streams,} and $\mathbf{F}_j\in\mathbb{C}^{M_j \times D_j}$ is the complete linear precoder matrix of BS $j$ which is the horizontal concatenation of UEs' linear precoders.

\textcolor{black}{The post-processed signal of UE $k$ after performing receive beamforming is given by}
\begin{equation}\label{y_tilde_k}
\tilde{\mathbf{y}}_k = \sum_{j\in \mathcal{J}}\mathbf{W}_k^* \mathbf{H}_{k,j}\mathbf{x}_j + \mathbf{W}_k^*\mathbf{z}_k
\end{equation}
%(including both small- and large-scale fading components)
where $\mathbf{W}_k\in\mathbb{C}^{N_k \times n_k}$ is the linear combiner (receive beamforming matrix) of UE $k$.
The presented signal model is applicable for all types of transmit beamforming and receive combining.
In MIMO mmWave systems, hybrid (analog and digital) beamforming can be implemented to reduces cost and power consumption of large antenna arrays \cite{SS}. 
In this paper, we employ the SVD beamforming technique to obtain the beamforming matrices at the transmitters and receivers \cite{TWC}.

\subsection{Interference Signal and Power}
In the omnidirectional interference model (OIM), the instantaneous rate of UE $k$ when connected to BS $j$ can be computed as
\begin{equation}\label{R_kj_Full}
R_{k,j}^\text{OIM} = \log_2\Big |\mathbf{I}_{N_k} + \frac{P_j}{Q_j}(\mathbf{V}^\text{OIM}_{k,j})^{-1}\mathbf{H}_{k,j} \mathbf{H}_{k,j}^*\Big |
\end{equation}
where $\mathbf{H}_{k,j}$ is the mmWave channel between BS $j$ and UE $k$, $Q_j$ is the quota of BS $j$, and $\mathbf{V}_{k,j}$ is the interference and noise covariance matrix given as
\begin{align}\label{V_No_BF}
\mathbf{V}_{k,j}^\text{OIM}&=\Big( \sum_{i=1}^J \sum_{l=1,l\neq k}^K \frac{P_i}{Q_i}\mathbf{H}_{l,i}\mathbf{H}_{l,i}^*\Big)+N_0 \mathbf{I}_{N_k}
\end{align}
In the case of beamforming transmission, when UE $k$ is connected to BS $j$, its instantaneous rate can be obtained as
\begin{equation}\label{R_kj_Actual}
R_{k,j}^\text{BIM}= \log_2\Big |\mathbf{I}_{n_k} + \frac{P_j}{Q_j}(\mathbf{V}^\text{BIM}_{k,j})^{-1}\mathbf{W}_{k}^*\mathbf{H}_{k,j}\mathbf{F}_{k,j} \mathbf{F}_{k,j}^* \mathbf{H}_{k,j}^*\mathbf{W}_{k}\Big |
\end{equation}
where $\mathbf{V}_{k,j}$ is the interference and noise covariance matrix given as
\begin{align}\label{V_BF}
&\mathbf{V}_{k,j}^\text{BIM}= \mathbf{W}_{k}^*\mathbf{H}_{k,j}\Big( \sum_{\substack{l\in \mathcal{Q}_{j} \\ l\neq k}} \frac{P_j}{Q_j}\mathbf{F}_{l,j} \mathbf{F}_{l,j}^* \Big ) \mathbf{H}_{k,j}^*\mathbf{W}_{k}  \nonumber \\
&+ \mathbf{W}_{k}^* \Big( \sum_{\substack{i\in \mathcal{J} \\ i\neq j}} \sum_{\substack{l\in \mathcal{Q}_i}} \frac{P_i}{Q_i}\mathbf{H}_{k,i}\mathbf{F}_{l,i} \mathbf{F}_{l,i}^* \mathbf{H}_{k,i}^* \Big ) \mathbf{W}_{k} + N_0 \mathbf{W}_k^*\mathbf{W}_k
\end{align}
where $\mathbf{F}_{k,j}\in\mathbb{C}^{M_j\times n_k}$ is the linear precoder (transmit beamforming matrix) for each UE $k$ associated with BS $j$ which separates user data streams, $\mathbf{W}_k\in\mathbb{C}^{N_k \times n_k}$ is the linear combiner (receive beamforming matrix) of UE $k$, and $N_0$ is the noise power spectral density. 
%The presented signal model is applicable for all types of transmit beamforming and receive combining. 

For both transmission types, the interference and noise covariance matrix $\mathbf{V}_{k,j}$ is of significant interest. The diagonal values of this matrix represent the interference plus noise power at each receiving antenna, and the off-diagonal values represent the correlation between the interference at different pairs of antennas. In this paper, we are mainly interested in the interference plus noise power, i.e. the diagonal values of $\mathbf{V}_{k,j}$. Specifically, we will examine the total interference power across all receiving antennas after receive combiner at a typical interfered receiver. This interference power depends on the association between UEs and BSs as discussed in the next section.

%%%%%%%%%%%%%%%%%%%%%%%%%%%%%%%%%%%%%%%%%%%%%%%%%%%%%%%%%%%%%%%%%%%%%%%%%%%%%%%
%%%%%%%%%%%%%%%%%%%%%%%%%%%%%%%%%%%%%%%%%%%%%%%%%%%%%%%%%%%%%%%%%%%%%%%%%%%%%%%

\section{Load Balancing and max-SINR User Association}\label{LBUA_prob}
%Association and Activation Matrices
We follow the mmWave-specific user association model proposed in \cite{TWC} which takes into account the dependency between user association and interference structure in the network. This model is suitable for mmWave systems where the channels are probabilistic and fast time-varying, and the interference depends on the highly directional connections between UEs and BSs.
In this model, during each time slot $t$ the instantaneous CSI remains unchanged such that we can implement per-time-slot unique association. 
Note that the network operator can also choose to implement the proposed user association algorithms per multiple time slots, based on the channel CSI in the first time slot or the averaged channel CSI. Such a choice will lead to a trade-off between user association overhead and resulting network performance. All analysis and results in this paper are for per-time-slot association.

\vspace*{-1em}
\subsection{User Association via Activation Vector}
We follow the problem formulation given in \cite{TWC}. In this formulation the \textit{activation vector} $\mathbold{\beta}(t)$ is defined as
\begin{equation}\label{Beta_eq}
\mathbold{\beta}(t)=[\beta_1(t), ..., \beta_K(t)]^T
\end{equation}
where $\beta_k(t)$ is the called the \textit{activation factor} of UE $k$ and represents the index of BS to whom user $k$ is associated with during time slot $t$, i.e., $\beta_k(t)\in\mathcal{J}$ with $k\in\mathcal{K}$.  
Considering these definitions, the relationship between the activation set of BS $j$ and the activation factors can be described as
\begin{equation}\label{Q_j}
\mathcal{Q}_j(t) = \{ k: \beta_k(t)=j\}.
\end{equation}
We assume each UE can be associated with only one BS (unique association) at any time slot $t$, i.e.,
\begin{equation}
\mathcal{Q}_j(t) \cap \mathcal{Q}_i(t) = \varnothing,~~j\neq i \;,
\quad \quad 
\label{union}
\bigcup\limits_{j=1}^{J}\mathcal{Q}_j(t) = \mathcal{K}
\end{equation}
where (\ref{union}) indicates during each time slot all UEs are served by the BSs. In this paper, we employ max-SINR and load balancing user association schemes to study the interference model of an mmWave cellular network. 
\subsection{Max-SINR User Association}
% \textbf{From TWC paper:} 
Max-SINR user association is a traditional user association scheme which utilizes the highest SINR or largest received power to associate an UE with a BS \cite{3GPP2013}.
This technique has been working well in LTE cellular networks
at microwave frequencies where all cells are usually homogeneous
macro cells. The emergence of heterogeneous networks (HetNets) with smaller, lower
power pico, femto and relay BSs requires a different look at
user association to ensure load balancing across the network,
avoiding the case that all UEs connect to only a macro BS 
because of its strongest SINR and overload it. 
Max-SINR has been augmented in HetNets with techniques such as cell breathing \cite{CellBreath} and cell biasing \cite{CellBias} which can actually or artificially change the transmit power of a BS to attract or repel more UEs. These techniques, however, cannot precisely control the load on each BS.

\subsection{Load Balancing User Association}
For load balancing user association, the activation factors must satisfy the following load constraints
\begin{align}
\sum_{j\in\mathcal{J}} 1_{\beta_k(t)}(j) &\leq 1, ~~\forall k\in \mathcal{K}\label{TFA_cons_1}\\
\sum_{k\in\mathcal{K}}1_{\beta_k(t)}(j). n_k &\leq D_j, ~~\forall j\in \mathcal{J}\label{TFA_cons_2}
\end{align}
where $1_{\beta_k(t)}(j)=1$ if $\beta_k(t)=j$, and $1_{\beta_k(t)}(j)=0$ if $\beta_k(t)\neq j$.
The constraints in (\ref{TFA_cons_1}) reflect the fact that each UE cannot be associated with more than one BS in each time slot (unique association), and the resource allocation constraints in (\ref{TFA_cons_2}) denote that the total number of data streams of UEs served by each BS cannot exceed the total number of available data streams on that BS. We assumed that user association is performed during each time slot $t$, and thus we dropped the time index $t$ for notational simplicity in the analysis in Sec. III.

While load balancing user associations have been studied for sub-6 GHz LTE systems \cite{Andrews,Caire}, few works have considered mmWave networks. A load balancing user association algorithm, called worst connection swapping (WCS), for mmWave-enabled HetNets is introduced  in \cite{TWC} to maximize a network utility function including the throughput. This algorithm takes into account the dependency between user association and interference structure of the network, and is based on the intuition that swapping the worst UE-BS connection is likely to provide the UE a stronger link to another BS and/or reduce the interference, which consequently improves the UE's transmission rate. 

%\begin{figure}
%\centering
%%\includegraphics[scale=.3]{PDF_AverageRate2.pdf}
%\includegraphics[scale=.35]{PDF_user_rates.pdf}
%\caption{PDF of the average spectral efficiency per UE under (a) max-SINR user association, (b) load balancing WCS user association}
%\end{figure}

%%%%%%%%%%%%%%%%%%%%%%%%%%%%%%%%%%%%%%%%%%%%%%%%%%%%%%%%%%%%%%%%%%%%%%%%%%%%%%%
%%%%%%%%%%%%%%%%%%%%%%%%%%%%%%%%%%%%%%%%%%%%%%%%%%%%%%%%%%%%%%%%%%%%%%%%%%%%%%%
% MATLAB Files:
% 1. Generate channels using the following m-file
% 2. /Users/alireza/Desktop/PhD Stuff/CVX iMac/M117_COMCAS_Ch_Gen.m
% 3. Result of Step 2: save('M117_COMCAS_mmWave_channels_10Ch_50Loc.mat')
% 4. /Users/alireza/Desktop/PhD Stuff/CVX iMac/M117_COMCAS_Main2.m
% 5. Result of Step4: load('Sim_for_M117_COMCAS_10Locs_50CHs_4BS_16UE.mat')     % Sep. 8, 2019
\section{Numerical Results}\label{NumRes}
In this section, we evaluate the performance of the proposed BIM in the downlink of an mmWave cellular network with $J=4$ BSs and $K=16$ UEs operating at 28 GHz over a bandwidth of 1 GHz. 
\textcolor{black}{The mmWave channels are generated using the ACM and NYUSIM as described in Sec. II, and each channel (between a BS and a UE) is composed of 4 clusters with 7 subpaths per cluster.} In order to implement 3D beamforming, each BS is equipped with a UPA of size $8\times 8$ ($M_j=64, j\in\mathcal{J}$), and each UE is equipped with three antenna panels of size $4\times 1$ ($N_k=4, k\in\mathcal{K}$) designed for mmWave band, where only one panel (based on UE orientation and hand grip) is active at any time. The number of data streams (layers) intended for each UE is $n_k=4, k\in\mathcal{K}$. The transmit power of each BS is 30 dBm and the noise power spectral density is $-174$ dBm/Hz.
Network nodes are deployed in a $300 \times 300~\textrm{m}^2$ square where the BSs are placed at the center of $150 \times 150~\textrm{m}^2$ subsquares, and the UEs are distributed randomly according to a uniform distribution. 
We compare the performance of BIM with OIM considered in the literature \cite{Andrews, Caire} and show that the proposed BIM has a much better performance in terms of  user data rates and interference power at users.

%The Monte Carlo simulations are performed over 50 different locations and 10 channel realizations per location. 
\begin{figure}
\centering
\includegraphics[scale=.35]{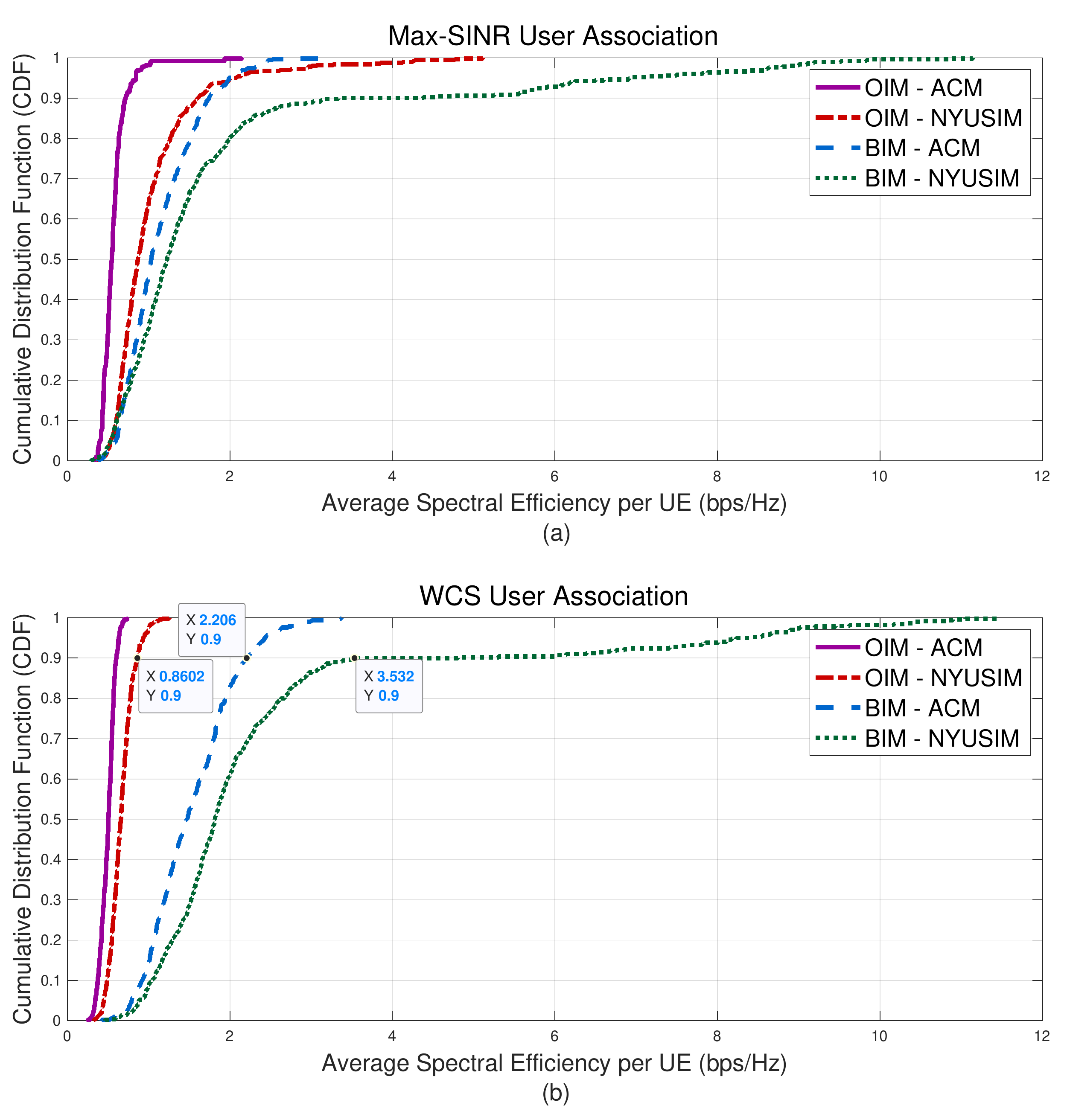}
\caption{CDF of the average spectral efficiency per UE under (a) max-SINR user association, (b) load balancing WCS user association}
\label{CDF_SE_perUE}
\end{figure}

Fig. \ref{CDF_SE_perUE} shows the CDF of average spectral efficiency per UE. As we can see from this figure, BIM, which employs SVD beamforming technique, results in a higher spectral efficiency under both Max-SINR association and WCS association. \textcolor{black}{For example, at 90\% level, the simulations using NYUSIM show that the BIM results in more than 3 times higher spectral efficiency compared to the OIM under WCS user association.
Moreover, this figure indicates that mmWave channels generated using measurement-based NYUSIM provide a better spectral efficiency than using ACM based on 3GPP channel model. This observation is in good agreement with the results obtained in \cite{NYUSIM, Rappaport_TVT}. For example, at 90\% level, NYUSIM results in about 60\% higher spectral efficiency than ACM under WCS user association (see Fig. \ref{CDF_SE_perUE}.b).} 
% (3.532-2.206)/2.206 = 60%

\begin{figure}
\centering
\includegraphics[scale=.35]{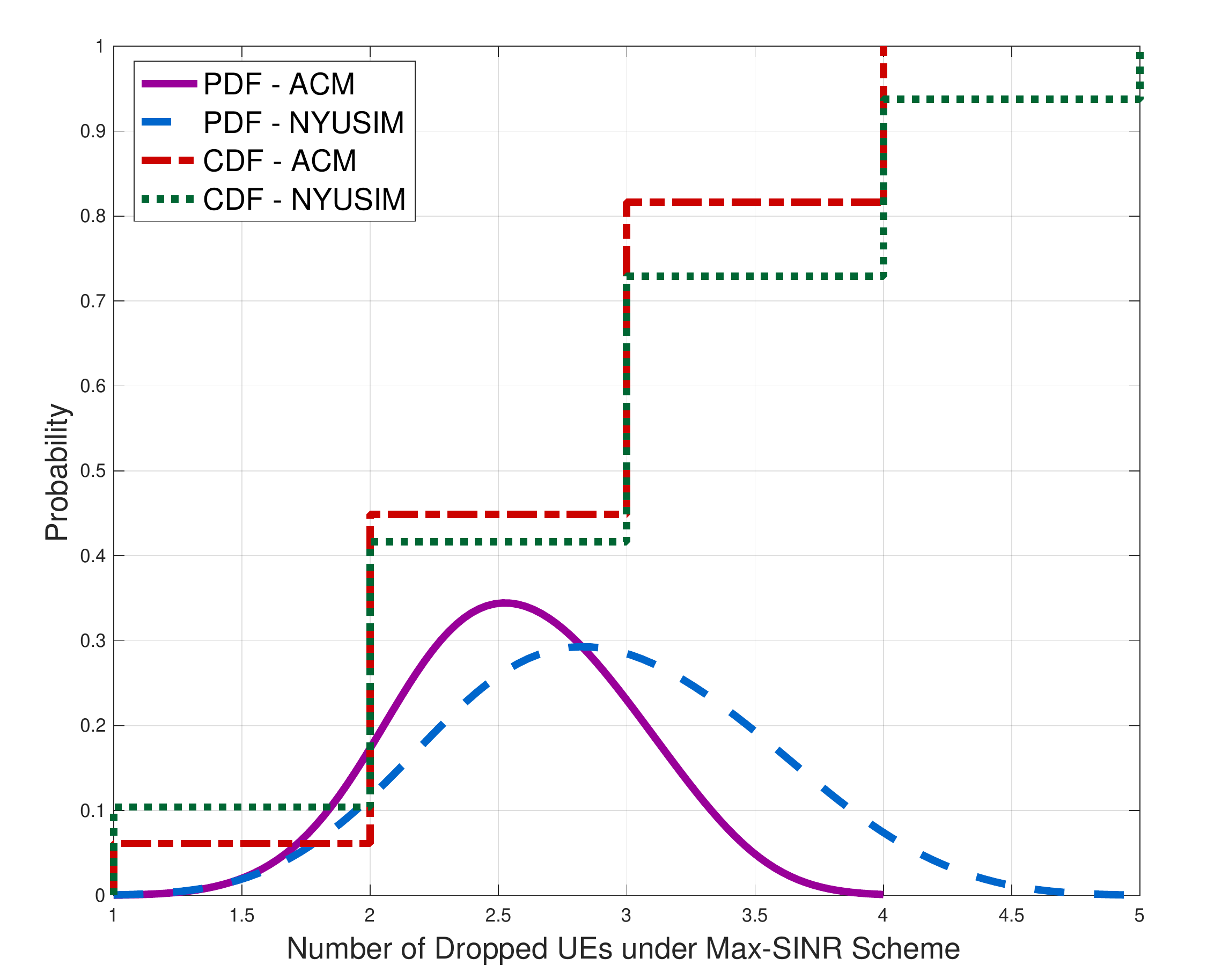}
\caption{PDF and CDF of the number of dropped UEs under max-SINR user association scheme}
\label{dropped_UEs}
\end{figure}

Furthermore, it can be inferred that the WCS algorithm outperforms the Max-SINR scheme when using BIM, while this is not the case for OIM. The reason is that the WCS user association algorithm is designed specifically for directional beamforming transmissions. These results show the importance of having a proper interference model while performing user association. Note that in max-SINR user association the overloading UEs are dropped, while the WCS algorithm considers the BSs' loads and equally distributes the UEs among the BSs ($Q_j=4, j\in\mathcal{J}$). Fig. \ref{dropped_UEs} depicts the PDF and CDF of the number of dropped UEs under max-SINR user association scheme. The CDF curves show that the probability of having at least 2 dropped UEs is about 94\% using ACM and 90\% using NYUSIM.

\begin{figure}
\centering
\includegraphics[scale=.35]{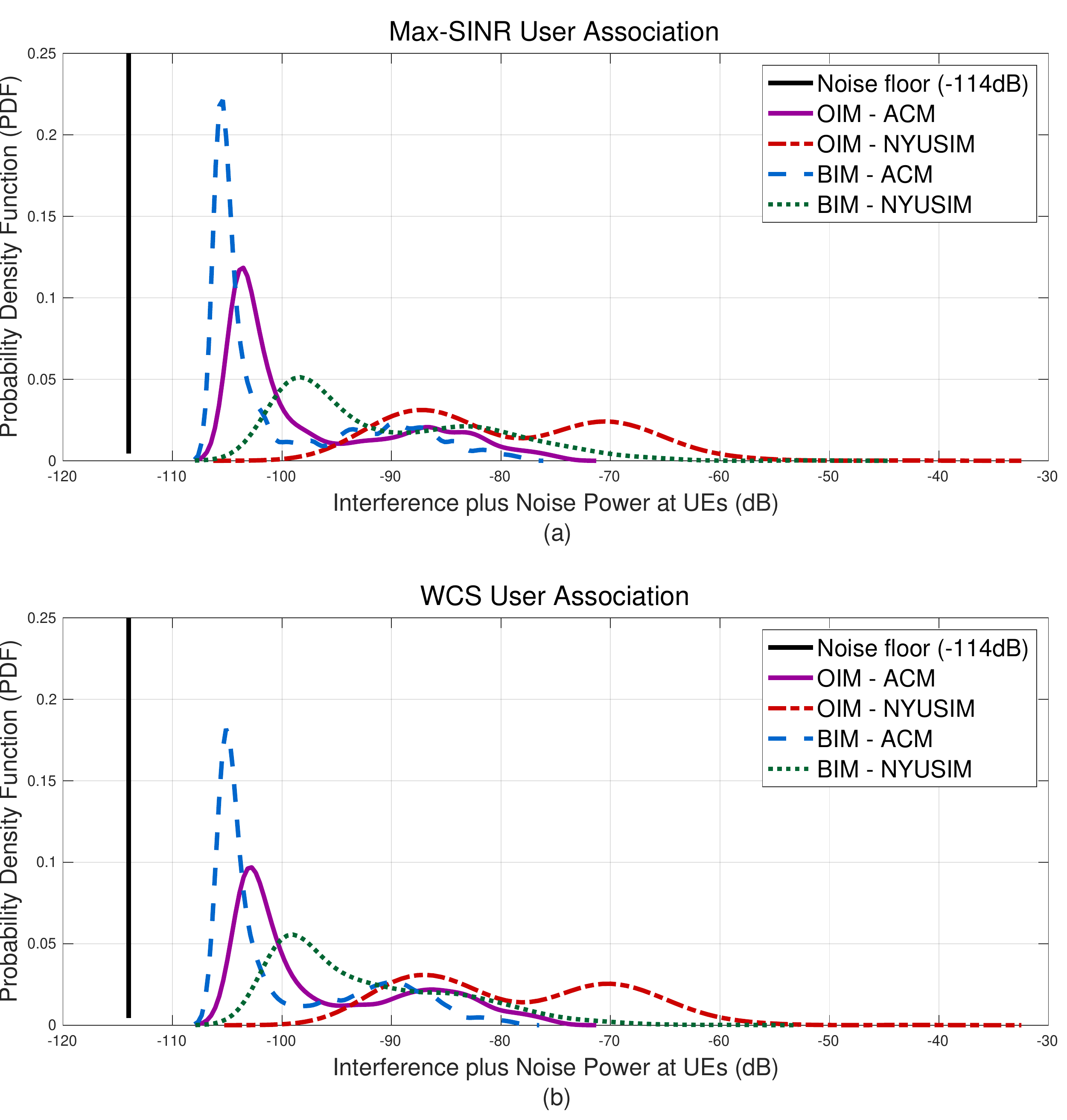}
\caption{PDF of the interference plus noise power at UEs under (a) max-SINR user association, (b) load balancing WCS user association}
\label{PDF_Interf}
\end{figure}

\begin{figure}
\centering
\includegraphics[scale=.35]{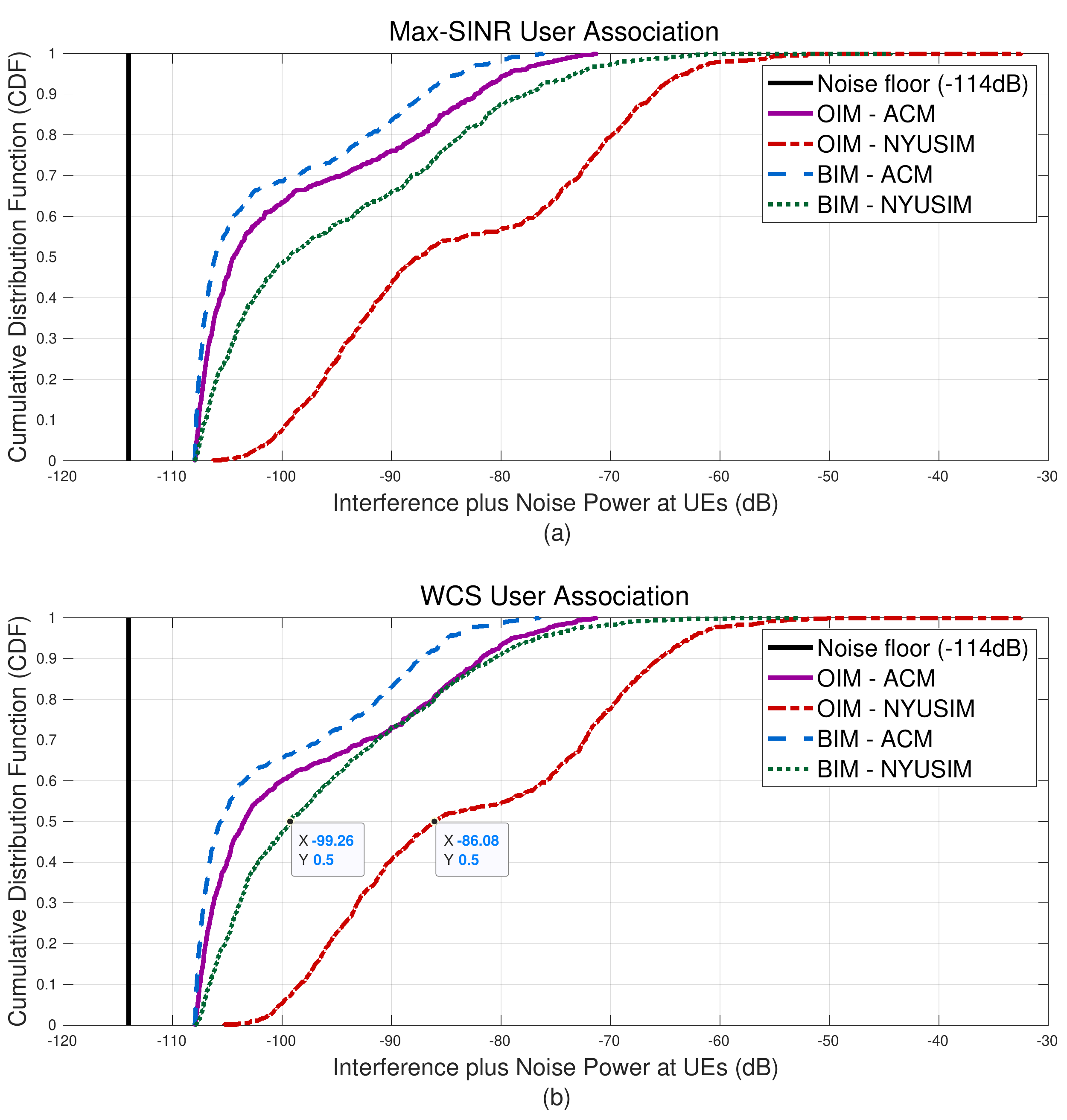}
\caption{CDF of interference plus noise power at UEs under (a) max-SINR user association, (b) load balancing WCS user association}
\label{CDF_Interf}
\end{figure}

Figs. \ref{PDF_Interf} and \ref{CDF_Interf} depict the PDF and CDF of the interference plus noise power at UEs in the network, respectively. 
The interference plus noise power for OIM and BIM are the trace of interference and noise covariance matrices in (\ref{V_No_BF}) and (\ref{V_BF}), respectively. 
The vertical line shows the noise floor which is -114 dB for the network bandwidth of 1 GHz. It can be seen that the minimum interference power is -108 dB which is higher than the noise level, making the network interference-limited rather than noise-limited. 
%The 6 dB difference is due to the fact that the terms $N_0\mathbf{I}_{N_k}$ in (\ref{V_No_BF}) and $N_0\mathbf{W}_k^*\mathbf{W}_k$ in (\ref{V_BF}) are both $4 \times 4$ matrices.
As shown in Fig. \ref{CDF_Interf}, the interference plus noise power is much lower when using directional beamforming transmissions compare to omnidirectional transmissions. 
The figures show that BIM has a superior performance compared to OIM by having the interference plus noise power more concentrated at lower values.
\textcolor{black}{As an example, at 50\% level, the interference plus noise power using the BIM is about 15\% lower than that for OIM (see Fig. \ref{CDF_Interf}.b).}
These results also confirm that even under beamforming transmissions, there is still a considerable amount of interference in an mmWave network. 
%Also, it can be inferred that beamforming transmissions can effectively mitigate the effect of interference and improve the spectral efficiency of users as shown in Figs 1 and 2. 

\section{Conclusion}
We examined the interference power distributions in an mmWave network using beamforming transmission under two user association schemes, and contrast with those under omnidirectional transmission. Numerical results show that beamforming can effectively mitigate network interference and hence significantly enhances the average spectral efficiency of UEs, also user association can considerably alter network interference and throughput structures. 
We further observed that the measurement-based NYUSIM channel generator results in higher spectral efficiency than the 3GPP-based ACM. Thus it is important to consider the multiple factors inter-playing when characterizing network interference and throughput in an mmWave system.

\bibliographystyle{IEEEtran}
\bibliography{References}

\begin{thebibliography}{10}
\providecommand{\url}[1]{#1}
\csname url@samestyle\endcsname
\providecommand{\newblock}{\relax}
\providecommand{\bibinfo}[2]{#2}
\providecommand{\BIBentrySTDinterwordspacing}{\spaceskip=0pt\relax}
\providecommand{\BIBentryALTinterwordstretchfactor}{4}
\providecommand{\BIBentryALTinterwordspacing}{\spaceskip=\fontdimen2\font plus
\BIBentryALTinterwordstretchfactor\fontdimen3\font minus
  \fontdimen4\font\relax}
\providecommand{\BIBforeignlanguage}[2]{{%
\expandafter\ifx\csname l@#1\endcsname\relax
\typeout{** WARNING: IEEEtran.bst: No hyphenation pattern has been}%
\typeout{** loaded for the language `#1'. Using the pattern for}%
\typeout{** the default language instead.}%
\else
\language=\csname l@#1\endcsname
\fi
#2}}
\providecommand{\BIBdecl}{\relax}
\BIBdecl

\bibitem{Andrews}
Q.~Ye, B.~Rong, Y.~Chen, M.~Al-Shalash, C.~Caramanis, and J.~G. Andrews, ``User
  association for load balancing in heterogeneous cellular networks,''
  \emph{IEEE Trans. Wireless Commun.}, vol.~12, no.~6, pp. 2706--2716, 2013.

\bibitem{Caire}
D.~Bethanabhotla, O.~Y. Bursalioglu, H.~C. Papadopoulos, and G.~Caire,
  ``{Optimal user-cell association for massive MIMO wireless networks},''
  \emph{IEEE Trans. Wireless Commun.}, vol.~15, no.~3, pp. 1835--1850, 2016.

\bibitem{60GHz}
G.~Athanasiou, P.~C. Weeraddana, C.~Fischione, and L.~Tassiulas, ``Optimizing
  client association for load balancing and fairness in millimeter-wave
  wireless networks,'' \emph{IEEE/ACM Trans. Netw.}, vol.~23, no.~3, pp.
  836--850, 2015.

\bibitem{TWC}
A.~Alizadeh and M.~Vu, ``Load balancing user association in millimeter wave
  mimo networks,'' \emph{IEEE Trans. Wireless Commun.}, vol.~18, no.~6, pp.
  2932--2945, 2019.

\bibitem{Niknam}
S.~{Niknam} and B.~{Natarajan}, ``On the regimes in millimeter wave networks:
  Noise-limited or interference-limited?'' in \emph{2018 IEEE International
  Conference on Communications Workshops (ICC Workshops)}, May 2018, pp. 1--6.

\bibitem{SS}
{3GPP TR 38.901 v15.0.0}, ``{Study on channel model for frequencies from 0.5 to
  100 GHz (Release 15)},'' Tech. Rep., Jun. 2018.

\bibitem{Nokia}
T.~A. Thomas, H.~C. Nguyen, G.~R. MacCartney, and T.~S. Rappaport, ``{3D mmWave
  channel model proposal},'' in \emph{Proc. IEEE 80th Veh. Technol. Conf. (VTC
  Fall)}, 2014, pp. 1--6.

\bibitem{NYUSIM}
S.~{Sun}, G.~R. {MacCartney}, and T.~S. {Rappaport}, ``A novel millimeter-wave
  channel simulator and applications for {5G} wireless communications,'' in
  \emph{2017 IEEE International Conference on Communications (ICC)}, May 2017,
  pp. 1--7.

\bibitem{Rappaport_TVT}
S.~{Sun}, T.~S. {Rappaport}, M.~{Shafi}, P.~{Tang}, J.~{Zhang}, and P.~J.
  {Smith}, ``Propagation models and performance evaluation for {5G}
  millimeter-wave bands,'' \emph{IEEE Trans. Veh. Technol.}, vol.~67, no.~9,
  pp. 8422--8439, 2018.

\bibitem{RapLetter}
M.~K. Samimi, T.~S. Rappaport, and G.~R. MacCartney, ``Probabilistic
  omnidirectional path loss models for millimeter-wave outdoor
  communications,'' \emph{IEEE Wireless Commun. Lett.}, vol.~4, no.~4, pp.
  357--360, 2015.

\bibitem{HetNetC-RAN}
M.~Peng, Y.~Li, J.~Jiang, J.~Li, and C.~Wang, ``Heterogeneous cloud radio
  access networks: a new perspective for enhancing spectral and energy
  efficiencies,'' \emph{IEEE Wireless Communications}, vol.~21, no.~6, pp.
  126--135, 2014.

\bibitem{samimi20163}
M.~K. Samimi and T.~S. Rappaport, ``{3-D} millimeter-wave statistical channel
  model for {5G} wireless system design,'' \emph{IEEE Trans. Microw. Theory
  Techn.}, vol.~64, no.~7, pp. 2207--2225, 2016.

\bibitem{3GPP2013}
{3GPP TR 36.872 v12.1.0}, ``{Small cell enhancements for E-UTRA and E-UTRAN -
  Physical layer aspects},'' Tech. Rep., Dec. 2013.

\bibitem{CellBreath}
Y.~Bejerano and S.-J. Han, ``Cell breathing techniques for load balancing in
  wireless lans,'' \emph{IEEE Trans. Mobile Comput.}, vol.~8, no.~6, pp.
  735--749, 2009.

\bibitem{CellBias}
A.~Damnjanovic, J.~Montojo, Y.~Wei, T.~Ji, T.~Luo, M.~Vajapeyam, T.~Yoo,
  O.~Song, and D.~Malladi, ``A survey on {3GPP} heterogeneous networks,''
  \emph{IEEE Wireless Commun.}, vol.~18, no.~3, 2011.

\end{thebibliography}


%\begin{comments}

%%\begin{IEEEbiography}{Alireza Alizadehl}
%%Biography text here.
%%\end{IEEEbiography}

%%\begin{IEEEbiographynophoto}{John Doe}
%%Biography text here.
%%\end{IEEEbiographynophoto}

%\newpage

%%\begin{IEEEbiographynophoto}{Jane Doe}
%%Biography text here.
%%\end{IEEEbiographynophoto}

%\end{comments}

%\vfill


\end{document}